\documentclass[12pt]{elsart}

\usepackage{amsmath}
\usepackage{amsfonts}
\usepackage{graphicx}
\def\R{{\bf R}}
\begin{document}

\begin{frontmatter}

\title{Complexity invariance by replication in the quantum square well}

\author[rlr]{Ricardo L\'{o}pez-Ruiz} and
\ead{rilopez@unizar.es}
\author[jsr]{Jaime Sa\~{n}udo}
\ead{jsr@unex.es}

\address[rlr]{
DIIS and BIFI, Facultad de Ciencias, \\
Universidad de Zaragoza, E-50009 Zaragoza, Spain}
\address[jsr]{
Departamento de F\'isica, Facultad de Ciencias, \\
Universidad de Extremadura, E-06071 Badajoz, Spain}

%\date{\today}

\begin{abstract}
A new kind of invariance by replication of a statistical measure of complexity is considered.
We show that the set of energy eigenstates of the quantum infinite square well displays
this particular invariance. Then, this system presents a constant complexity for all the
energy eigenstates.
\end{abstract}


\end{frontmatter}

\maketitle

Quantum systems can be interpreted as objects evolving in an space of distributions.
Different scalar magnitudes, such as energy and statistical complexity, can be calculated 
on them. The energy has a clear physical meaning \cite{landau1981}. 
Among other applications, it can be used to find the equilibrium states of a system.
In this same way, it has been also shown that the complexity can give us 
some insight about the equilibrium configuration in some quantum systems. 
For instance, the ground state of the $H_2^+$ molecule
has been studied under the optics of complexity measures \cite{sen2008}. 
In this case, Montgomery and Sen have reported that the minimum of the statistical complexity
\cite{lopez1995} as a function of the internuclear distance for this molecule gives an 
accurate result comparable with that obtained with the minimization of the energy.
This fact could suggest that energy and complexity are two magnitudes strongly related
for any quantum system. But this is not the general case. See, for example, the behavior of
both magnitudes in the H-atom \cite{sanudo2008} and in the quantum isotropic harmonic 
oscillator \cite{sanudo2008+}. In both systems, the degeneration of the energy 
is split by the statistical complexity, in such a way that the minimum of complexity 
for each level of energy is taken on the wave function with the maximum orbital angular momentum.
This seems to mean that energy and complexity are two independent variables and, hence, 
it is not possible to estimate the value of one of them knowing the other one.

In this work, it is our aim to continue unveiling the independence of energy and complexity. 
Thus, we study the possibility of having a quantum system where both magnitudes
present a contrary behavior to that found in the H-atom and in the quantum isotropic 
harmonic oscillator for those magnitudes. That is, we wonder
if there exists such a system where degeneration 
of the complexity can be split by the energy. 
Our answer will be affirmative. We proceed to show it in two steps.
First, we establish a new type of invariance by
replication for the statistical complexity, and, second, we show that the energy eingestates of the 
quantum infinite square well fulfill the requirements of this kind of invariance. Therefore,
the degeneration of complexity in this quantum system is revealed to be broken by the energy.

Let us start by recalling the definition of a measure of complexity $C$ \cite{lopez1995,lopez2002},
the so-called $LMC$ complexity, that is defined as
\begin{equation}
C = H\cdot D\;,
\end{equation}
where $H$ represents the information content of the system and $D$ gives an idea
of how much concentrated is its spatial distribution. 
For our purpose, we take a version used in Ref. \cite{lopez2002}
as quantifier of $H$. This is the simple exponential Shannon entropy \cite{dembo1991},
that takes the form, 
\begin{equation}
H = e^{S}\;,
\end{equation}
where $S$ is the Shannon information entropy \cite{shannon1948},
\begin{equation}
S = -\int p(x)\;\log p(x)\; dx \;,
\label{eq1}
\end{equation}
with $x$ representing the continuum of the system states and $p(x)$
the probability density associated to all those states.
We keep for the disequilibrium the form originally introduced in 
Refs. \cite{lopez1995,lopez2002}, that is,
\begin{equation}
D = \int p^2(x)\; dx\;.
\label{eq2} 
\end{equation}

Lloyd and Pagels \cite{lloyd1988} recommend that a complexity measure
should remain essentially unchanged  under replication. Different types
of replication can be defined on a given probability density.
One of them was established in Ref. \cite{lopez2002}.
Here, we present a similar kind of replication, in such manner that
the complexity $C$ of $m$ replicas of a given distribution is equal to the
complexity of the original one. 

Thus, if $\R$ represents the support of the density function $p(x)$,
with $\int_{\R} p(x)\,dx = 1$, take $n$ copies $p_m(x)$,
$m=1,\cdots,n$, of $p(x)$,
\begin{equation}
p_m(x) = \; p(n (x-\lambda_m))\, ,\;\;
1\leq m\leq n \, ,
\label{eqrep}
\end{equation}
where the supports of all the $p_m(x)$, centered at $\lambda_m's$
points, $m=1,\cdots,n$, are all disjoint. Observe that
$\int_{\R} p_m(x)\, dx = \frac{1}{n}$, what makes the replicas union
\begin{equation}
q_n(x)=\sum_{i=1}^n p_m (x)
\label{eqqn}
\end{equation}
to be also a normalized probability distribution,
$\int_{\R} q_n(x)\, dx = 1$. For every $p_m(x)$, a
straightforward calculation shows that
\begin{equation}
S(p_m) =  \frac{1}{n}\, S(p) \, ,
\end{equation}
\begin{equation}
D(p_m) =  \frac{1}{n}\, D(p) \, .
\end{equation}

Taking into account that the $m$ replicas are supported on
disjoint intervals on \R, we obtain
\begin{equation}
S(q_n)  =  S(p) \, , \\
\end{equation}
\begin{equation}
D(q_n)  =  \, D(p) \, .
\end{equation}
Then,
\begin{equation}
C(q_n) = C (p) \, ,
\end{equation}
and this type of invariance by replication for $C$ is shown.

Let us see now that the probability density of the eigenstates of the energy
in the quantum infinite square well display this type of invariance.
The wave functions representing these states for a particle in a box, that is 
confined in the one-dimensional interval $[0,L]$, are given by \cite{cohen1977}
\begin{equation}
\varphi_k(x)=\sqrt{2\over L}\,\sin\left(k\pi x\over L\right),\,\, k=1,2,\ldots
\end{equation}
Taking $p(x)$ of expressions (\ref{eq1}) and (\ref{eq2}) as the probability density 
of the fundamental state $(k=1)$,
\begin{equation}
p(x)=\mid \varphi_1(x)\mid^2, 
\end{equation}
we can interpret the probability density of the $kth$ excited state, 
\begin{equation}
q_k(x)=\mid \varphi_k(x)\mid^2, 
\end{equation}
as the union of $k$ replicas of the fundamental state density, $p(x)$, in the $k$ disjoint intervals 
$[{(m-1)L/ k}, {mL/ k}]$, with $m=1,2,\ldots,k$. That is, we find expression (\ref{eqqn}), 
$q_k(x)=\sum_{i=1}^k p_m (x)$, with 
\begin{equation}
p_m(x)={2\over L}\,\sin^2\left({k\pi x\over L}-\pi(m-1)\right),\,\, m=1,2,\ldots,k,
\end{equation}
where in this case the $\lambda_m's$ of expression (\ref{eqrep})
are taken as $(m-1)L/k$. Therefore, we conclude that the complexity is degenerated
for all the energy eigenstates of the quantum infinite square well. Its value can be
exactly calculated. Considering that $L$ is the natural length unit in this problem, 
we obtain
\begin{equation}
C(p)=C(q_k)={3\over e}=1.1036...
\end{equation}
In the general case of a particle in a $d$-dimensional box of width $L$ in each dimension,
it can be also verified that complexity is degenerated for all its energy eigenstates with a 
constant value given by $C=(3/e)^d$. 

We conclude by remarking that in the same way that the complexity breaks the energy degeneration
in the H-atom and in the quantum isotropic harmonic oscillator, we have shown here that
the contrary behavior is also possible. In particular, the complexity is constant
for the whole energy spectrum of the $d$-dimensional quantum infinite square well. 
This result can be interpreted as due to the same functional 
form displayed by all the energy eigenstates of this system. Therefore, we suggest that,
at the present level of interpretation of the statistical complexity, 
the study of this magnitude in a quantum system allows us to infer some properties 
on its structural conformation.

{\bf Acknowledgements}
The authors acknowledge some financial support by spanish grant DGICYT-FIS2005-06237.


\begin{thebibliography}{10}

\bibitem{landau1981} L.D. Landau and L.M. Lifshitz,
{\it Quantum Mechanics: Non-Relativistic Theory}, volume 3, 
Third Edition, Butterworth-Heinemann, Oxford, 1981.

\bibitem{sen2008} H.E. Montgomery Jr. and K.D. Sen,
Phys. Lett. A, {\bf 372} (2008) 2271.    

\bibitem{lopez1995} R. Lopez-Ruiz, H.L. Mancini, and X. Calbet,
Phys. Lett. A {\bf 209} (1995) 321.

\bibitem{sanudo2008} J. Sa\~{n}udo and R. Lopez-Ruiz,
Phys. Lett. A {\bf 372} (2008) 5283.

\bibitem{sanudo2008+} J. Sa\~{n}udo and R. Lopez-Ruiz,
J. Phys. A: Math. Theor. {\bf 41} (2008) 265303.

\bibitem{lopez2002} R.G. Catalan, J. Garay, and R. Lopez-Ruiz,
Phys. Rev. E {\bf 66} (2002) 011102.

\bibitem{dembo1991} A. Dembo, T.A. Cover, and J.A. Thomas,
IEEE Trans. Inf. Theory {37} (1991) 1501.

\bibitem{shannon1948} C.E. Shannon,
{\it A mathematical theory of communication},
Bell. Sys. Tech. J. {\bf 27} (1948) 379; ibid. (1948) 623.

\bibitem{lloyd1988} S. Lloyd and H. Pagels, Ann. Phys. (N.Y.) 
{\bf 188} (1988) 186.

\bibitem{cohen1977} C. Cohen-Tannoudji, B. Diu and F. Laloe,
{\it Quantum Mechanics}, 2 vols., Wiley, New York, 1977.

\end{thebibliography}
\end{document}